\documentstyle[12pt,aasms4]{article}

\def\feka{Fe K$\alpha$}
\def\chandra{{\it Chandra}} 
 
\def\asca{{\it ASCA}} 
\def\hst{{\it HST}}

\def\rosat{{\it ROSAT}} 
\def\ein{{\it Einstein}} 
 
\def\merlin{{\it MERLIN}} 
\def\vla{{\it VLA}} 
\def\vlba{{\it VLBA}} 
\def\sirtf{{\it SIRTF}} 
\def\lum{erg s$^{-1}$}
\def\flux{erg cm$^{-2}$ s$^{-1}$}
\def\nh{cm$^{-2}$}
\def\arcsec{$^{\prime\prime}$}

\def\ltsima{$\; \buildrel < \over \sim \;$}
\def\simlt{\lower.5ex\hbox{\ltsima}} 
\def\gtsima{$\; \buildrel > \over \sim \;$}
\def\simgt{\lower.5ex\hbox{\gtsima}} 

\def\cl{\centerline}

\input{psfig.tex} 

\begin{document}
\title{A survey of extended radio jets in AGN with Chandra and HST:
First Results} 

\author{Rita M. Sambruna} 
\affil{George Mason University, Dept. of Physics and Astronomy and School of
Computational Sciences, MS 3F3, 4400 University Drive, Fairfax, VA 22030
(rms@physics.gmu.edu)} 

\author{L. Maraschi and F. Tavecchio}
\affil{Osservatorio Astronomico di Brera, via Brera 28, 20121 Milano, Italy} 

\author{C. Megan Urry}
\affil{Yale University, New Haven, CT 06520}

\author{C. C.  Cheung}
\affil{MS 057, Department of Physics, Brandeis University, Waltham, MA
02454}

\author{G. Chartas}
\affil{The Pennsylvania State University, Department of Astronomy and
Astrophysics, 525 Davey Lab, State College, PA 16802}

\author{R. Scarpa} 
\affil{ESO, Chile} 

\author{Jessica K. Gambill} 
\affil{George Mason University, School of
Computational Sciences, 4400 University Drive, Fairfax, VA 22030}

\begin{abstract}

We present the first results 
from an X-ray and optical survey of a sample of AGN radio jets with
\chandra\ and \hst. We focus here on the first six sources observed at
X-rays, in four of which
a bright X-ray jet was detected for the first time. In three out of
four cases optical emission from the jet is also detected in our \hst\
images. We compare the X-ray morphology with the radio as derived
from improved processing of archival \vla\ data and we construct
spectral energy distributions (SED) for the most conspicuous emission
knots. In most cases the SEDs, together with the similarity of the
X-ray and radio morphologies, favor an inverse Compton origin of the
X-rays. The most likely origin of the seed photons is the Cosmic
Microwave Background, implying the jets are still relativistic on
kiloparsec scales. However, in the first knot of the PKS 1136-135 jet,
X-rays are likely produced via the synchrotron process. In all four
cases bulk Lorentz factors of a few are required. The radio maps of
the two jets not detected by either \chandra\ or \hst\ suggest that
they are less beamed at large scales than the other four detected
sources. Our results demonstrate that, at the sensitivity and
resolution of \chandra, X-ray emission from extragalactic jets is
common, yielding essential information on their physical properties.

{\sl Subject Headings:}{Galaxies: active --- galaxies: jets ---
(galaxies:) quasars: individual --- X-rays: galaxies}

\end{abstract} 

\section{Introduction}

Jets are a common feature of radio-loud Active Galactic Nuclei (AGN),
providing a means for powering the luminous giant radio lobes which
first marked extreme activity in these systems. A detailed
understanding of jet physical conditions (e.g., emission mechanisms,
balance between particle and magnetic field energies, acceleration
processes) requires knowing their morphologies and spectral energy
distributions at various energies, hence the importance of
multiwavelength imaging. However, while hundreds of kpc-scale radio
jets are known (Bridle \& Perley 1984), until now only a handful were
detected in the optical with \hst\footnote[1]{Based on observations
made with the NASA/ESA {\it Hubble Space Telescope}, obtained at the
Space Telescope Science Institute, which is operated by the
Association of Universities for Research in Astronomy, Inc., under
NASA contract NAS5--26555.} and ground-based telescopes (Sparks et
al. 1994; Scarpa et al. 1999). Even fewer jets were known at X-ray
wavelengths from \ein\ and \rosat, at limited resolution and
sensitivity (Feigelson et al. 1981; Harris \& Stern 1987; Biretta et
al. 1991). These studies established the non-thermal nature of the
multiwavelength emission of jets, attributing the smooth
radio-to-optical continuum to synchrotron emission, while the X-rays
were interpreted either as the high-energy tail of the synchrotron
emission, or as due to inverse Compton scattering of the synchrotron
photon themselves off the high energy electrons in the jet, the
synchrotron-self Compton process (see Harris 2001 for a recent
review). 
 
The advent of the \chandra\ X-ray Observatory is opening a new chapter
in the study of jets. The main features that make \chandra\ ideal to
study X-ray emission from kpc-scale jets in AGN are its high angular
resolution (0.5\arcsec\ FWHM at 1 keV) and sensitivity in the 0.2--10
keV energy band. Indeed, several jets were already detected at X-rays
with \chandra, a few of which were previously known from the optical
(Chartas et al. 2001; Hardcastle et al. 2001; Pesce et al. 2001;
Wilson et al. 2001; Worrall et al. 2001).  The first surprise came
from the \chandra\ ``first light'' image of the distant blazar PKS
0637--752, where a bright X-ray jet was discovered with a morphology
close to the radio but with only a weak optical counterpart (Chartas
et al. 2000; Schwartz et al. 2000).  Canonical synchrotron or SSC
models fail to reproduce adequately the radio-to-X-ray spectral
distribution of the jet. The latter is instead well explained by a
model invoking inverse Compton scattering of the cosmic microwave
background photons (IC/CMB) off the jet electrons (Tavecchio et
al. 2000; Celotti et al. 2001).  The attractiveness of this model is
that it accounts successfully for the shape of the X-ray spectrum of
PKS 0637--752, and yields minimum estimates for the jet kinetic
power. The IC/CMB model was applied successfully to explain the X-ray
emission of other jets (Chartas et al. 2001; Sambruna et al. 2001).

The detection of the PKS 0637--752 jet and its unexpected properties
raised several questions. In particular: is PKS 0637--752 an
exceptional object or are X-ray and optical emission a common feature
of extended radio jets in AGN? What is the role of the IC mechanism as
opposed to the synchrotron mechanism in accounting for the large scale
X-ray emission of jets? 

To address these questions, we obtained \chandra\ (cycle 2) and \hst\
(cycle 9) time to perform a survey of 17 carefully selected radio
jets, to find their X-ray and optical counterparts and start
addressing their physical properties. Given the sensitivity of both
\chandra\ and \hst, and the selection criteria of our jet sample (see
below), this survey represents an unbiased probe of the occurrence of
X-ray and optical emission from extragalactic jets in AGN. We also
have a program to image these jets uniformly at radio wavelengths
using the \vla, \vlba, and \merlin\ (Cheung et al. 2002).

In this paper, we present the results of the first six objects
observed with \chandra. We focus on the X-ray counterparts of the
extended radio features, leaving the discussion of the nuclear
properties to a future paper (Gambill et al. 2002). We find a high
detection rate of the jets at X-rays, with four out of six objects
showing a bright X-ray jet with complex and different
morphologies. Three out of these four X-ray jets also have optical
counterparts in our \hst\ images. While the \hst\ results will be
fully presented in a future paper (Scarpa et al. 2002), here we use
optical fluxes extracted from the \hst\ images. The remaining jets of
the survey will be presented in a future publication.

The plan of this paper is as follows. In \S~2 we review the sample
selection criteria, the observations and the data analysis, in \S~3 we
present the results for the six \chandra\ targets, including a
comparison with the radio data, and in \S\S~4 and 5 we discuss the
implications. Throughout this work, H$_0=75$ km s$^{-1}$ Mpc$^{-1}$
and $q_0=0.5$ are adopted. Spectral indices are defined as $F_{\nu}
\propto \nu^{-\alpha}$. 

\section{Observations} 

\subsection{Sample Selection}

The targets of the program were selected from the radio, without any
{\it a priori} knowledge of their optical and X-ray emission
properties. In this sense our survey is unbiased toward detections at
shorter wavelengths.   

Specifically, the survey sample was extracted from the list of known
radio jets of Bridle \& Perley (1984) and Liu \& Xie (1992), according
to the following criteria: 1) The radio jet is \simgt 3\arcsec, i.e.,
long enough to be easily resolved with \chandra\ and \hst; 2) The
radio jet has radio surface brightness $S_{1.4~GHz}$ \simgt
5mJy/arcsec$^2$ at $>$ 3\arcsec\ from the nucleus, i.e., bright enough
to be detected in reasonable \chandra\ and \hst\ exposures for average
values of the radio-to-X-ray and radio-to-optical spectral indices,
$\alpha_{rx} \sim 0.8$ and $\alpha_{ro} \sim 0.8$; and 3)
High-resolution (1\arcsec\ or better) published radio maps show that
at least one bright (\simgt 5 mJy) radio knot is present at $>$
3\arcsec\ from the nucleus. These criteria gave a sample of 17 radio
jets, spanning a range of redshifts, core and extended radio powers,
and classification (quasar, BL Lac, radio galaxy). However, most
sources show one-sided jets, indicating substantial beaming.

The awarded exposures were 10 ks per target with \chandra\ ACIS-S
(occasionally, targets were observed for slightly more or less than 10
ks to accomodate gaps in the \chandra\ schedule) and one orbit per
target with \hst. The X-ray flux limit for a 2$\sigma$ detection in a
typical 10 ks ACIS-S observation is F$_{0.4-8~keV} \sim 2 \times
10^{-15}$ \flux. It is important to note that the short \chandra\ and
\hst\ exposures were designed to find the X-ray/optical counterpart of
the radio jet, but are not sufficient for a detailed study of their
morphologies and spectra. 

Here we report the results for the first six observed sources:
0723+679, 1055+018, 1136--135, 1150+497, 1354+195, and 2251+134. Their
basic properties, reported in Table 1, are representative of the whole
sample. A bright X-ray jet was detected in 0723+679, 1136--135,
1150+497, and 1354+195. Except for 0723+679, an optical counterpart to
the X-ray jet is present as well in the \hst\ images.

\subsection{Data Reduction and Analysis}

The \chandra\ observations were performed with ACIS-S, with the source
at the aimpoint of the S3 chip. As most sources have bright X-ray
cores, to reduce pileup of the nucleus (which can affect the study of
the jet inner regions) we used a 1/8 subarray mode with only the S3
chip turned on, giving an effective frametime of 0.4 s. Nevertheless,
the core X-ray flux of 1354+195, 1055+018, 1136--135, and 1150+497 was
still piled-up, as suggested by the observed count rates of the cores
(Gambill et al. 2002). The fraction of core pileup is \simgt 5\%,
according to the measured counts/frame (\simgt 0.27 c/frame) and
Figure 6.25 of the \chandra\ Proposers Observatory Guide (Rev. 3.0,
December 2000). In addition, for each source a range of roll angles
was specified in order to locate the jet away from the charge transfer
trail from the nucleus and avoid flux contamination. The \chandra\
data were reduced following standard screening criteria and using the
latest calibration files provided by the \chandra\ X-ray Center. Pixel
randomization was removed, and only events for \asca\ grades 0, 2--4,
and 6 and the energy range 0.4--8 keV, where the background is
negligible and the ACIS-S calibration best known, were retained. We
also checked that no flaring background events occurred during the
observations. After screening, the effective exposure times range
between 9--11 ks (Table 2).

Several knots were detected at X-rays in the \chandra\ images (Table
2). X-ray counts from each knot were extracted in a circular region of
radius 1\arcsec, with the background estimated in a circular region of
radius 8\arcsec\ at a nearby position free of serendipitous X-ray
sources. The source extraction radius of 1\arcsec\ encircles \gtsima
90\% of the flux at 1 keV, based on the ACIS-S encircled energy
fraction (Fig. 6.3 in the \chandra\ Proposer Observatory Guide).

Let $S$ be the total (before background subtraction) counts in the
source area $A_S$, and $b$ the background counts in the area
$A_B$. The background counts rescaled to the source region are $B=b
\times A_S/A_B$. The net counts of the source are thus $N=S-B$.  The
uncertainties on the net X-ray counts, $\sigma_N$, were calculated
according to the formula 
$\sigma_N=\sqrt{(\sigma_S)^2 + (\sigma_B)^2}$, where $\sigma_S$ and
$\sigma_B$ are the uncertainties on the source and background,
respectively. The latter were calculated following Gehrels (1986), as
appropriate in a regime of low counts: $\sigma_S=1+\sqrt{S + 0.75}$
and $\sigma_B=(1+\sqrt{b + 0.75})\times(A_S/A_B)$. 

For all the radio features not detected in the \chandra\ images, we
estimated the 3$\sigma$ upper limits to the X-ray flux. These were
extracted in a 1\arcsec-radius region centered on the radio position
of the knot, with the background estimated as discussed above. In the
cases of the non-detected jets of 2251+134 and 1055+018, rather than
quoting upper limits for each individual radio feature, we give a
total upper limit integrated over the whole jet emission.

In the cases of 1150+497 and 1354+195, available radio images show the
presence of jet knots at $\sim$ 2\arcsec\ from the core. At this
distance, the \chandra\ PSF of the X-ray core is non-negligible, given
the bright core emission of the two sources (Fig. 3 and 4).
Inspection of the X-ray maps of 1150+497 and 1354+195, however, shows
enhancements of the flux above the PSF wings at the position of the
inner radio knots, suggesting possible X-ray emission of the
knots. These were denominated ``knots A'' in Table 2 for both
sources. To derive their net X-ray flux, we adopted the following
procedure. We extracted the counts from a circular aperture of radius
1\arcsec, centered on the radio position of the knot. To subtract the
contribution of the PSF, the background was evaluated in a circular
region with the same radius and at the same distance from the nucleus,
but at various azimuth angles.  We find that the net
(background-subtracted) counts vary by as much as 30\% depending on
the background region, suggesting local disuniformities of the core
PSF most likely due to the fact that significant pileup affects the
PSF in both sources. Given the large excursion of values, the counts
reported in Table 2 for knots A of 1150+497 and 1354+195 are simple
average values, while the uncertainties are their standard deviations.
Thus, they represent rather conservative uncertainties for the X-ray
fluxes of these knots.

We also analyzed archival radio images at 5 GHz from the {\it Very
Large Array}\footnote[2]{The National Radio Astronomy Observatory is a
facility of the National Science Foundation operated under cooperative
agreement by Associated Universities, Inc.} database (Thompson et
al. 1980). Data for 1055+018, 1136--135, and 1354+195 were retrieved
from the \vla\ archive at NRAO and calibrated in AIPS using standard
procedures. The data were taken from programs AM213 (17 Aug 1987),
AE59\footnote[3]{The flux scale of this experiment was based on
contemporaneous measurements of 0727--115 (4 Jy) obtained from the
University of Michigan on-line database (the average of three
measurements within one week of the archival VLA observation).} (30
Oct 1988), and AB331B (20 Apr 1985), respectively. To the best of our
knowledge, these data are unpublished.  The images of 0723+679 and
1150+497 were published originally by Owen \& Puschell (1984) and have
been reimaged by us using calibrated (u,v) data kindly provided by
F. Owen. The image of 2251+134 was reimaged from data originally
published in Price et al. (1993).  We chose to restore the radio
images of the X-ray detected knots using circular beams with
FWHM=0.86\arcsec\ in order to match the resolution of the final
smoothed Chandra images. The two images of the non-detected jets were
restored with circular beams with FWHM=1.5\arcsec\ for 1055+018 and
1.5\arcsec\ for 2251+134. More details will be given in Cheung et
al. (2002).

\hst\ observations were performed with \verb+STIS+ in \verb+CLEAR+
mode, with an effective wavelength of 5852 \AA. The calibration used
for flux extraction is based on the keyword \verb+PHOTFLAM+=$8.97
\times 10^{-20}$ erg s$^{-1}$ cm$^2$ \AA$^{-1}$ count$^{-1}$,
corresponding to 1.08 $\times 10^{-7}$ Jy/count s$^{-1}$. The \hst\
observations will be the subject of a separate paper (Scarpa et
al. 2002).

Upper limits to the optical flux for extended radio features
(counterlobes) in each object were calculated using measurements of
the background surface brightness in the corresponding \hst\
images. The surface brightness measurements were converted to fluxes
assuming a square aperture with an area of 1 arcsec$^{2}$. For
undetected optical counterparts to radio jet knots, we estimate
conservatively a 3$\sigma$ upper limit of 0.06 $\mu$Jy. This limit is
based on the fact that the weakest detected jet knots in our \hst\
images were 0.02 $\mu$Jy (one in 1150+497 and two in 1040+123; Scarpa
et al. 2002). Since the background signals in all our \hst\ images
were comparable, we took 0.02 $\mu$Jy as a conservative estimate for
the detection limit of any undetected knots for all the sources. The
expected/observed optical positions of most jet features in these
objects place them reasonably well away from contamination by light
from their host galaxies, except in the cases of 0723+679 and
2251+134. For selected features in these two objects (the counterlobe
and jet, respectively), we measured the Poissonian noise directly in
galaxy subtracted images within a circular aperture with radius
1\arcsec\ coincident with the expected positions of the features. We
expect the jet knots in 0723+679 to be compact enough that they would
have manifested themselves above the smooth galaxy profile, therefore,
the 3$\sigma$ upper limit of 0.06 $\mu$Jy was applied to them.

\section{Results} 

Figures 1--5 show the \chandra\ ACIS-S images of the six sources
(color scale). The detected X-ray jets are plotted in Fig. 1--4
while Fig. 5 shows the sources where the jets were not detected
(1055+018 and 2251+134). The X-ray images were produced by smoothing
the raw \chandra\ data with a Gaussian of width=0.3\arcsec\ in the
energy range 0.4--8 keV, with final resolution of 0.86\arcsec\
FWHM. Overlaied on the X-ray images are the radio contours from the
archival \vla\ data, smoothed to the same (for the detected X-ray
jets) resolution as \chandra.  In the four sources in Figure 1--4
(0723+679, 1136--135, 1150+497, and 1354+195), an X-ray counterpart to
the radio jet is clearly detected, with knotty and widely different
morphologies.  Their properties are given in Table 2: distance from
the core, Position Angle (PA), and net X-ray counts. We stress that
the nomenclature adopted in Table 2 is X-ray driven and does not
reflect the previous nomenclature of the radio knots.  The apparent
(projected) jet lengths vary from 6.7\arcsec\ (33 kpc) in 1136--135 to
27\arcsec\ (145 kpc) in 1354+195.  The jets are not resolved in a
direction perpendicular to their axes. A summary of the X-ray and
radio properties is given below for each detected jet.

\vspace{0.1in}

\noindent {\bf 0723+679}: A bright X-ray knot (B) is detected at
6.3\arcsec\ from the nucleus, with a weaker one (A, $1.6\sigma$
detection) at 4.3\arcsec. Both coincide with radio knots in the \vla\
radio map (Owen \& Puschell 1984).  The jet appears to bend after the
first knot. Faint X-ray emission is also visible from the hot spot of
the extended radio counter-lobe, with the peak of the X-ray emission
being displaced from the peak of the radio emission by $\sim
1$\arcsec. On the whole, the jet X-ray and radio brightnesses track
each other very well. There is no indication in the \hst\ image of an
optical counterpart to any of the extended radio or X-ray
features. The X-ray core exhibits interesting spectral properties,
including an \feka\ emission line (Gambill et al. 2002).

\noindent{\bf 1136--135}: Two X-ray knots are detected at 4.5\arcsec\
(A) and 6.7\arcsec\ (B) from the nucleus; both have a bright optical
counterpart in our \hst\ images. Knot A has an extremely weak radio
counterpart; knot B, the brightest in X-rays, has a relatively weak
radio counterpart (Saikia et al. 1989). Overall, after knot B, the
X-ray and radio brightnesses along the jet behave oppositely, with the
X-rays peaking at $\sim$ 7\arcsec\ from the core, in correspondence to
a secondary peak in the radio, while the radio emission peaks at
10\arcsec\ at knot C, for which we can set only an upper limit in
X-rays.

\noindent{\bf 1150+497}: The X-ray jet has at least three bright knots
at 2.1\arcsec\ (A), 4.3\arcsec\ (B), and 7.9\arcsec\ (C) from the
core.  The final knot coincides with a terminal hot spot in the radio
image. As shown by the PA in Table 2, the X-ray jet wiggles by $\Delta
PA \sim 20$ degrees, following closely the radio (Owen \& Puschell
1984). All the three X-ray knots have an optical counterpart in our
\hst\ images.

\noindent{\bf 1354+195}: This spectacular, well-collimated X-ray jet
is also the longest one in our sample, extending up to 27\arcsec\ (145
kpc projected length). It consists of a string of nine X-ray knots,
closely tracing the radio morphology (Murphy, Browne, \& Perley
1993). X-ray emission is also present from the southern radio hot spot
(knot I), which appears misaligned with respect to the jet axis by
$\Delta PA \sim 10$ degrees, and off-center by 2\arcsec\ with respect
to its radio counterpart. Only inner knots A and B have optical
counterparts in our \hst\ images. Knot I is off the \hst\ field of
view and its optical counterpart is not known. 

\vspace{0.1in}

Table 3 gives multiwavelength information for the jets, in the form of
radio (5 GHz), optical (5852 \AA), and X-ray (1 keV) fluxes for the
various knots, and the derived broad-band spectral indices:
radio-to-optical $\alpha_{ro}$, optical-to-X-ray $\alpha_{ox}$, and
radio-to-X-ray $\alpha_{rx}$. Fluxes at radio and optical wavelengths
were extracted using the same 1\arcsec\ circular regions as for the
X-rays. The X-ray fluxes were obtained from the the counts in Table 2
assuming a power law with photon index $\Gamma=2.0$ and absorbing
column density fixed to the Galactic value. No correction for
reddening was applied to the optical fluxes; however, this correction
is negligible compared to the large uncertainties of the optical
fluxes.
We also list 3$\sigma$ upper limits (see discussion above) for the
radio knots that were not detected at optical or/and X-rays, and for
the counterlobes (CL). 

\section{Interpretation} 

\subsection{General constraints from the radio and X-ray morphologies} 


X-rays from extended radio jets in AGN can derive from different
processes.  Non-thermal mechanisms, suggested by the presence of
relativistic electrons necessary to account for the radio emission,
include synchrotron radiation (from the high-energy end of the
electron population), or inverse Compton scattering processes, with
the seed photons provided by the synchrotron photons themselves (SSC)
or by external radiation fields. Recently, Tavecchio et al. (2000) and
Celotti et al. (2001) showed that if the plasma is still relativistic
on large (10 kpc or more) scales, the dominant source of seed photons
is the cosmic microwave background (CMB). In fact, the CMB radiation
density is amplified in the comoving frame of the emitting plasma by a
factor $\Gamma^2$, where $\Gamma$ is the bulk Lorentz factor of the
jet. The resulting X-ray emission is strongly beamed. As shown by
Dermer (1995; see also Brunetti et al. 2001) the emission cone of this
radiation is narrower than that of the synchrotron emission. In this
case the relative importance of the X-ray emission depends on the
viewing angle, being greater for small angles.

In this IC/CMB model, we expect the X-ray and radio morphologies of
the jet to be generally similar, as emission at both wavelengths is
produced by electrons in similar, relatively low, energy ranges
($\gamma_X \sim 100$, $\gamma_{\rm radio} \sim 1000$). However,
differences in the X-ray-to-radio brightness ratio can arise from, and
give information on, changes in magnetic field and/or bulk Lorentz
factor (deceleration) along the jet.

If X-rays are due to synchrotron emission, the radiating electrons
must have very high energies, a factor $\simeq 10^4$ larger than those
emitting in the radio and with correspondingly shorter lifetimes by
the same factor. Thus, electrons producing X-rays may cool rapidly
after the acceleration region, while radio electrons can survive and
diffuse in regions further away from the acceleration sites. The
expectation for the jet morphology is then that the X-ray emission
should peak before the radio. Inspection of Figure 1--4 shows that, in
the cases of 0723+679, 1150+497, and 1354+195, the X-ray and radio
morphologies are very close, suggesting that the X-ray emission from
these jets is due to IC/CMB. The fourth jet, 1136--135, is a complex
case. Knot A is bright at X-rays and in optical, but does not have a
strong radio counterpart. Knot B is bright at all three wavelengths,
while knot C at the end of the jet is bright only at the longer
wavelengths. This suggests that different emission processes for the
X-rays may dominate at different locations along the 1136--135 jet.

\subsection{Spectral Energy Distributions} 

Complementary and more easily quantifiable information is provided by
the spectral energy distributions (SED) of bright emission regions in
the jet.  These can be derived for a given knot extracting fluxes from
the same spatial region around the knot at the various wavelengths.

In interpreting the SED, a critical role is played by the optical
flux. Specifically, if the optical emission lies on the extrapolation
between the radio and X-ray fluxes or above it, the SED is compatible
with a single electron spectrum extending to high energies; instead,
if the optical emission falls well below the extrapolation, it argues
for different spectral components (and therefore different mechanisms
unless two electron populations are hypothesised) below and above the
optical range (e.g., synchrotron and IC respectively). Thus, in all
knots where IC/CMB dominates the X-ray emission, we expect an up-turn
of the spectrum in the X-ray band with respect to the radio-optical
extrapolation and thus an optical-to-X-ray index $\alpha_{ox}$ flatter
than the radio-to-optical index $\alpha_{ro}$.  Conversely, when
synchrotron dominates we expect $\alpha_{ro}$ \simlt $\alpha_{ox}$,
with the inequality holding when radiative losses are important in the
X-ray band.

Figure 6 shows the plot of the broad-band indices, $\alpha_{ro}$
versus $\alpha_{ox}$, from the data in Table 3. The dotted line marks
the locus of points for which $\alpha_{ro}=\alpha_{ox}$.  It can be
seen that most knots lie in the region $\alpha_{ro} > \alpha_{ox}$,
except for knot A in the jet of 1136--135 (Table 3). This behavior is
well consistent with the morphological properties discussed above.

To investigate more quantitatively the jet physical properties, we
computed synchrotron and IC/CMB emission models reproducing the SEDs
of the most conspicuous knots.  We note that with only three observed
fluxes the models are underconstrained. Following the procedure of
Tavecchio et al. (2000), we assume the flux extraction radius is the
size of the (spherical) emission region and compute for what values of
the magnetic field, $B$, and of the beaming factor,
$\delta$\footnote[4]{The beaming factor $\delta$ is here
defined as $\delta \equiv [\gamma(1-\beta \cos\theta)]^{-1}$, where
$\beta$ is the bulk velocity of the plasma in units of the speed of
light, $\gamma=(1-\beta^2)^{-1/2}$ the corresponding Lorentz factor,
and $\theta$ the angle between the velocity vector and the line of
sight.} it is possible to reproduce the radio and X-ray fluxes. As an
example, the results for 1150+497 (knot B) are shown in Figure 7a. The
dotted line represents the condition of energy equipartition between
high-energy electrons and magnetic field, as derived from the
synchrotron radio emission, which is the same for all models. The
dashed and continuous lines represent the appropriate values of $B$
and $\delta$ for the SSC and IC/CMB models, respectively. It is clear
from Figure 7a that SSC models require either no beaming or
substantial debeaming (implying a very high power of the jet) and come
close to equipartition only for very large values of $B$ and large
debeaming, while IC/CMB models meet the equipartition line at
reasonable values of $B$ and moderate values of $\delta$.  We thus
favor the latter and adopt equipartition as an additional constraint
to fix the model univocally.

In the cases where X-rays are attributed to direct synchrotron
emission, there is no ``independent'' constraint from the X-ray flux
and only the equipartition assumption survives.  In Fig. 7b we show the
synchrotron constraints for knot A of 1136--135. We also plot the
radiative lifetime of the X-ray emitting electrons as an additional
criterium (see below). 

Guided by the above parametric analysis, specific models were computed
to reproduce the SEDs of X-ray knots A and B of each detected jet with
synchrotron plus IC/CMB emission and assuming equipartition. The
results are shown in Figure 8 and the derived parameters, for the
electron spectrum, the beaming factor, $\delta$, and the magnetic
field, $B_{equip}$, are reported in Table 4. The upper and lower
limits of the electron spectrum, $\gamma_{max}$ and $\gamma_{min}$, 
determine the high energy cut-off of the synchrotron component and the
low energy cut-off of the IC/CMB component, respectively. These two
cut-offs are crucial in determining the SED in the intermediate
IR-optical-UV-soft X-ray ranges, a region presently not constrained by
the data. In all cases we used $\gamma_{min}=10$.

For three of the X-ray jets (0723+679, 1150+497, and 1354+195) the
favored X-ray emission process is IC/CMB. Optical emission can be
attributed either to synchrotron or to IC/CMB processes, depending on
the value of $\gamma_{min}$ and $\gamma_{max}$.  Clearly, only the
optical {\it spectrum} can constrain these two parameters and
discriminate between the synchrotron and IC/CMB models. Observations
with \hst\ at different wavelengths are needed to this purpose. 

In the case of knot A of 1150+497 and knots A and B of 1354+195, we
reproduce optical emission with the synchrotron component, while for
knot B in 1150+497 we assume that the dominant process is IC.
Consider the case of 0723+679. We have only an upper limit for the
optical emission which constrains the synchrotron component to cut off
before the optical range. Any optical emission at the level of the
present upper limit could be due either to the high-energy tail of the
synchrotron component (slightly higher $\gamma_{max}$) or to the
low-energy end of the IC component (slightly lower $\gamma_{min}$).



The case of 1136--135 is different. As discussed above, this jet
appears to have a ``mixed'' morphology, and indeed the SEDs of the two
brightest X-ray knots differ (Fig. 8). For knot A, where $\alpha_{ro}$
\simlt $\alpha_{ox}$, we suggest that synchrotron emission is a
plausible emission process. Still, the emitted radiation should be
beamed ($\delta \sim 7$ for $B_{equip} \sim 10~ \mu$Gauss; Table 4),
or the magnetic field would be implausibly high and/or the
relativistic plasma far from equipartition (Fig. 7b).  Assuming a
magnetic field of $B=10~\mu$Gauss, the cooling time of the electrons
in the knot is $\sim 10^{11}$ s. This is consistent with the
light-crossing time for knot A, assuming its radius is 1\arcsec.
Thus, high-energy electrons can cool before they reach knot B, in
agreement with the SED. Note that the associated IC/CMB component is
only a factor $\simeq 3$ below the synchrotron one.

For knot B in the same jet, clearly $\alpha_{ro}$ \simgt $\alpha_{ox}$
and IC/CMB should be the main emission process.  The X-rays fading
further out and the increasing radio brightness should then be
attributed to a deceleration of the the relativistic plasma and to a
compression of the magnetic field, as expected at the outer boundary
of jets (e.g., G{\'o}mez 2001).

In summary, for the knots of the 0723+679, 1150+497, and 1354+195
jets, the IC/CMB scenario is the favored process for the X-ray
emission, both from the comparison of the radio and X-ray morphologies
and a more quantitative analysis of the SEDs. The model is consistent
with equipartition for moderate values of the beaming factor ($\delta
\sim 5-6$), with $B_{equip} \sim (2-10) \times 10^{-5}$ G. The SSC
model requires in general conditions very far from equipartition and
larger intrinsic powers in the jet ($P_j>10^{49}$ \lum). On the other
hands, the 1136--135 jet shows that various emission processes may
dominate the X-ray production at different locations in the jet.  This
could be due to a different maximum energy to which the electrons are
accelerated.


\subsection{Jet Power} 

The last two columns of Table 4 list the jet intrinsic power,
calculated assuming the parameters from the models, and the total
radiated jet power, i.e., the observed power integrated over the whole
solid angle (e.g., Sikora et al. 1997). Comparison of the two powers
shows that only a small fraction of the jet kinetic power is
dissipated into radiation. This implies small radiative efficiencies
and that most of the energy extracted from the central black hole is
stored in the bulk motion of the plasma. 

There have been suggestions (e.g., Ghisellini \& Celotti 2001) that
jet radiative efficiency decreases with increasing jet
luminosity. While based on a very limited sample, it is interesting to
note that, with the exception of 1136--135, the X-ray detected jets
lie at the higher-luminosity end of the distribution of jet radio
luminosity in Table 1. Clearly, confirmation of this result awaits
completion of our survey.

\subsection{The non-detections} 

Two radio jets of the survey, 1055+018 and 2251+134, were not detected
at either X-rays and optical (Fig. 5). Since their radio fluxes are
comparable to those of the 4 detected jets, there must be some
intrinsic differences responsible for their low X-ray emission.

A possible explanation could simply be an overall steeper electron
spectrum. This idea could be easily verified by acquiring
multifrquency radio observations.  Alternatively, their large scale
jets may be non-relativistic or seen at larger angles. Note that
indeed the radio maps do not appear strongly asymmetric and the upper
limits to their jet-to-counterjet ratios are smaller than in the other
four cases (Table 1). However, it is also true that their
core-to-extended radio flux ratio is very high, suggesting that the
core is beamed. It is tempting to suggest that in these objects the
jet decelerates not far from the core so that it is relativistic on pc
scale but non-relativistic on larger scales. In this case, since a
large part of the jet power must be dissipated in the inner part of
the jet, one naturally expects that the outer jet is faint,
contributing to enhance the value of the core-to-extended radio flux
ratio. High-resolution, multi-epoch radio observations could in
principle verify this scenario by providing a measure of the apparent
bulk speed in the jet inner regions.

\section{Discussion} 

As stressed above, the IC/CMB model works only if the plasma still has
bulk relativistic motion ($\Gamma >> 1$) on kpc scales and the viewing
angle is small ($\theta$ \simlt $\Gamma^{-1}$). Independent evidence
for large bulk velocities on kpc scales in the jets of our sample
where the IC/CMB process works at X-rays is provided by the radio, for
which the 4 X-ray detected jets have especially large ratios of the
jet-to-counterjet flux (Table 1). Some previous jet studies with
ultra-deep \vla\ observations have provided statistical evidence that
jets generally decelerate on kpc scales, with $\delta \sim$ a few
(Wardle \& Aaron 1997). Our results, however, are not necessarily in
disagreement, because our selection criteria biases us toward the more
beamed jets and, as Table 4 shows, only moderate beaming is required
from our models.  A picture where jets have a fast ``spine'' which
produces the X-ray radiation, and slow walls from which the
longer-wavelength radiation originates, as advocated by many authors
(e.g., Sol, Pelletier, \& Asseo 1989; Laing 1993; Owen et al. 1997;
Wardle \& Aaron 1997; Chiaberge et al. 2000), may be more
realistic. While more complex emission models should then be used, the
present data would be insufficient to constrain them.

Another possible scenario for the X-ray production from jets (Celotti
et al. 2001) is IC scattering of the beamed emission from a blazar
nucleus, which according to unification models is harbored by every
radio-loud AGN (Urry \& Padovani 1995). However, this process is
likely to be unimportant if the jet plasma is still relativistic on
large scales. In fact, in this case, the beamed photons from the inner
jet appear redshifted in the plasma rest-frame, thus energetically
unimportant.

On the other hand, IC scattering of the nuclear radiation could be
important for the X-ray emission from radio lobes where the plasma
reaches slower terminal velocities (Brunetti et al. 1997). This
process could possibly explain the X-ray emission from the counterlobe
of 0723+679, where a marginal detection is obtained from the short
\chandra\ exposure (Table 2 and Figure 1). In this model, since the
up-scattered photons are beamed back toward the nucleus, only the
counterlobe (i.e., the lobe opposite along the line of sight) is
enhanced, as is indeed the case in 0723+679.

An alternative possibility for the production of X-rays from
large-scale jets is the class of hadronic models, where the X-rays are
direct synchrotron radiation of ultra-relativistic protons (Aharonian
2001). This model predicts a smooth optical-to-X-ray continuum
spectrum if the protons are responsible for the longer wavelengths as
well, which can be tested with future UV observations. It also
requires much higher magnetic fields ($\sim$ mGauss) and much more
efficient acceleration processes than in the leptonic models.

In conclusion, analysis of the SEDs of the various jet knots yields
results which are consistent with the indications inferred from the
radio/X-ray morphologies. Thus, the shape of the SEDs and the
multiwavelength jet morphologies are powerful indicators of the
physical mechanisms operating in the jet. Importantly, different
processes may predominate at various locations in the jet for the
production of X-rays, depending on the local physical conditions. 

It is also worth remarking a few caveats affecting our analysis.  First,
the limited signal-to-noise ratio at both X-ray and optical wavelengths
gives room to alternative interpretations of the SEDs. Second, a variety
of conditions may exist within the relatively large extraction regions we
used (1\arcsec, dictated by the \chandra\ resolution), for example if the
emitting particle distributions are stratified or multiple shocks
exist. While higher angular resolutions at X-rays await future
generations of space-based telescopes, deeper follow-up X-ray and optical
observations of the new jets of this survey with \chandra\ and \hst\ can
at least remedy the first limitation of our analysis, in providing higher
quality images, smaller uncertainties on the fluxes, and X-ray and
optical continuum spectra for individual knots. Note that optical
observations are necessary to identify the mechanism responsible for the
X-ray emission. An additional important constraint will be provided by
future IR observations with \sirtf, probing a poorly known region in the
SEDs where the synchrotron peak is located.

\section{Conclusions} 

In summary, we presented the detection of four new X-ray extragalactic
jets with \chandra\ from our X-ray/optical survey, strengthening the
evidence that X-ray emission from the kpc-scales of these structures
is common. The new X-ray detections of jets imply the presence of
relativistic bulk motion on very large scales, at tens to hundreds of
kpc from the central black hole, setting new challenges for
theoretical models. X-ray emission is produced in localized regions,
likely shocks associated with the deceleration of the
jet. Furthermore, our data show that different emission processes for
the X-rays may dominate in the same jet at different locations,
depending on the local physical conditions (e.g., the maximum electron
energy, $\gamma_{max}$).  Future deeper images at both X-ray and
multiple optical frequencies will be essential to explore in greater
detail the energetics and physical properties of the jets.

\acknowledgements 

This project is funded by NASA grant GO1-2110A and grant
HST-GO-08881.01-A, from the Space Telescope Science Institute, which
is operated by AURA, Inc., under NASA contract NAS~5-26555. CMU
acknowledges further support from NASA grant NAG5-9327.  Radio
Astronomy at Brandeis University is supported by the National Science
Foundation. CCC thanks Dave Roberts and John Wardle for continued
support of his research. We are grateful to Frazer Owen for making his
calibrated \vla\ data available to us. This research has made use of
data from the University of Michigan Radio Astronomy Observatory which
is supported by funds from the University of Michigan. This work was
partly supported by the Italian Space Agency (ASI) under grants I/R
037/01 and I/R 105/00.

\newpage 

\cl{\bf Figure Captions}

\noindent Figure 1: \chandra\ ACIS-S image in the 0.4--8 keV energy
range of the newly discovered X-ray jet of 0723+679 (colors).
Overlaied are the radio contours from archival \vla\ data. Both the
colors and the contours are plotted logarithmically, in steps of
factor 2. The \chandra\ image is smoothed with a Gaussian of width
$\sigma$=0.3\arcsec, yielding a resolution of 0.86\arcsec\ FWHM.  The
radio image was restored with a circular beam with
FWHM=0.86\arcsec. The base level for the radio contours is 0.6
mJy/beam. In the image, 1\arcsec\ corresponds to 5.5 kpc.

\noindent Figure 2: Same as for Figure 1, but for 1136--135. The base level for
the radio contours is 0.8 mJy/beam. In the image, 1\arcsec\ corresponds to
4.9 kpc.

\noindent Figure 3: Same as for Figure 1, but for 1150+497. The base level for
the radio contours is 0.7 mJy/beam. In the image, 1\arcsec\
corresponds to 3.9 kpc.

\noindent Figure 4: Same as for Figure 1, but for 1354+195. The base level for
the radio contours is 0.9 mJy/beam. In the image, 1\arcsec\
corresponds to 5.4 kpc.

\noindent Figure 5: \chandra\ ACIS-S images in the 0.4--8 keV range of 1055+018
and 2251+134, for which no X-ray or optical jet was detected.
Overlaied are the radio contours from archival \vla\ data. Both the
colors and the contours are plotted logarithmically, in steps of
factor 2. The \chandra\ images were smoothed with a Gaussian of width
$\sigma$=0.3\arcsec, yielding a resolution of 0.86\arcsec\ FWHM. The
radio maps were restored with a circular beam with FWHM=1.5\arcsec\ for
1055+018 and 0.86\arcsec\ for 2251+134. The base level is 0.5 mJy/beam
and 1 mJy/beam for 1055+018 and 2251+134, respectively. In the images,
1\arcsec\ corresponds to 5.6 kpc for 1055+018 and 5.3 kpc for
2251+134.

\noindent Figure 6: Radio-to-optical index, $\alpha_{ro}$, versus the
optical-to-X-ray index, $\alpha_{ox}$, for individual knots in the
four newly detected X-ray jets. The dotted line marks the division
between knots where inverse Compton dominates over synchrotron for the
production of X-rays (see text). Most X-ray knots are unlikely to be
due to synchrotron emission from the high-energy tail of the
radio-emitting electron distribution.

\noindent Figure 7: {\it (a)} Allowed values of magnetic field and Doppler
beaming factor for knot B of 1150+497. {\it Solid line:} allowed
values if the X-rays come from inverse Compton scattering of cosmic
microwave background photons (the IC/CMB model). {\it Dashed line:}
allowed values of $B$ and $\delta$ if the observed X-ray emission is
produced by a synchrotron self-Compton model. {\it Dotted line:}
allowed values of $B$ and $\delta$ under the assumption of
equipartition between radiating particles and magnetic field assuming
that the low energy cut-off of the electron distribution is
$\gamma_{\rm min}=10$. The IC/CMB emission can clearly be consistent
with equipartition assumption for moderately large values of the
Doppler factor ($\delta \sim 6-7$). {\it (b)} Allowed values of
magnetic field and Doppler beaming factor for knot A of 1136--135 from
the synchrotron model. {\it Solid line and left vertical axis}: Values
of the Doppler factor $\delta $ as a function of the magnetic field
$B$ assuming equipartition between emitting electrons and magnetic
field. {\it Dashed line and right vertical axis:} Cooling time for
synchrotron emission of electrons producing 1 keV photons for
different values of the magnetic field and assuming the equipartition
value for $\delta $.

\noindent Figure 8: Radio-to-X-ray Spectral Energy Distributions (SEDs) for the
brightest X-ray knots in the four new X-ray jets of Figure 1--4. In
all panels, the left-handed vertical axes are in units of observed
flux, $\nu F_{\nu}$, while the right-handed axes are in units of
luminosity, $\nu L_{\nu}$.  Typical uncertainties on the fluxes are
33\% or larger. The X-ray flux is always above the extrapolation from
the radio-to-optical continuum, except for knot A in 1136--135,
suggesting a general dominance of inverse Compton for the production
of X-rays. In knot A of 1136--135, synchrotron is likely the dominant
process. The solid lines are the best-fit models (sum of all
components), with the parameters reported in Table 4.

\clearpage 


\scriptsize
\begin{center}
\begin{tabular}{llrlrrrcc}
\multicolumn{9}{l}{{\bf Table 1: The Sample}} \\
\multicolumn{9}{l}{   } \\ \hline
& & & & & & & & \\
Source & Alt name& $z$ & Type & logP$_{core}$ & logP$_{jet}$ & $R_i$& $J$ & N$_H^{Gal}$ \\ 
& & & & & & & & \\
(1) & (2) & (3) & (4) & (5) & (6) & (7) & (8) & (9) \\ 
& & & & & & & & \\ \hline
& & & & & & & & \\
0723+679 & 3C~179 & 0.846 & FSRQ & 33.88 & 33.56 & 0.5 & $>$ 21 & 4.31 \\
& & & & & & & & \\
1055+018 & 4C~+01.28 & 0.888 & FSRQ/BL & 34.76 & 33.15 & 16.5 & $>$ 3 & 3.40 \\
& & & & & & & & \\
1136--135& PKS & 0.554 & FSRQ & 33.50 & 33.46 & 0.30 &  $>$ 10 & 3.59 \\
& & & & & & & & \\
1150+497 &4C~+49.22& 0.334 & FSRQ & 33.04 & 32.39 & 2.30 &  $>$ 16 & 2.05 \\
& & & & & & & & \\
1354+195 &4C~+19.44& 0.720 & FSRQ & 34.35 & 33.43 & 2.80 & $>$ 56 & 2.18 \\
& & & & & & & & \\
2251+134 &4C~+13.85& 0.673 & QSO & 33.81 & 33.29 & 1.13 &  $>$ 14 & 4.98 \\
& & & & & & & & \\ \hline 

\end{tabular}
\end{center}

\noindent 
{\bf Explanation of Columns:} 1=Source IAU name; 2=Alternate name;
3=Redshift; 4=Optical/radio classification of core. FSRQ: Flat Spectrum
Radio Quasar; BL: BL Lac; QSO: Lobe-dominated Quasar; 
5=Log of the core power at 5 GHz (in erg s$^{-1}$ Hz$^{-1}$);
6=Log of the jet power at 5 GHz (in erg s$^{-1}$ Hz$^{-1}$);
7=Ratio of core to extended radio power at 5 GHz, corrected for the
redshift (observed value divided by (1+z)); 
8=Ratio of the jet-to-counterjet radio flux (see text for references);
9=Galactic column density in $10^{20}$ \nh. 


\newpage 


\scriptsize
\begin{center}
\begin{tabular}{lccrcc}
\multicolumn{6}{l}{{\bf Table 2: Chandra observations}} \\
\multicolumn{6}{l}{   } \\ \hline
& & & & & \\
Source & Exposure & Knot & Distance & PA & Counts \\ 
& & & & & \\
(1) & (2) & (3) & (4) & (5) & (6) \\
& & & & & \\ \hline
& & & & & \\
0723+679 & 9334 & A & 4.3 & 263 & 6 $\pm$ 4 \\
         &      & B & 6.3 & 274 & 18 $\pm$ 5 \\
         &      & CL$^a$ & 8.0 & 75 & 4 $\pm$ 3 \\ 
& & & & & \\
1055+018 & 9314 & $\cdots$ & $\cdots$ & $\cdots$ & $\cdots$ \\
&   &          &          &          &          \\
& & & & & \\
1136--135& 8906 & A & 4.5 & 295 & 22 $\pm$ 6 \\
         &      & B & 6.7 & 295 & 59 $\pm$ 9 \\
& & & & & \\
1150+497 & 9294 & A & 2.1 & 210 & 72 $\pm$ 22$^b$ \\
         &      & B & 4.3 & 190 & 27 $\pm$ 6 \\
         &      & C & 7.9 & 195 & 17 $\pm$ 5 \\
& & & & & \\
1354+195 & 9056 & A & 1.7 & 162 & 135 $\pm$ 40$^b$ \\
         &       & B & 3.6 & 160 & 10 $\pm$ 4 \\
         &       & C & 6.2 & 167 & 6 $\pm$ 4 \\
         &       & D & 8.8 & 168 & 4 $\pm$ 3  \\
         &       & E & 10.2 & 165 & 4 $\pm$ 3 \\
         &       & F & 13.2 & 165 & 12 $\pm$ 5 \\
         &       & G & 14.4 & 165 & 12 $\pm$ 5 \\
         &       & H & 16.8 & 163 & 7 $\pm$ 4 \\
         &       & I & 26.9 & 170 & 11 $\pm$ 4 \\
&& & & & \\
2251+134 & 9186 & $\cdots$& $\cdots$& $\cdots$& $\cdots$ \\
         & &         &         &         &        \\ 
& & & & & \\ \hline

\end{tabular}
\end{center}

\scriptsize 
\noindent 
{\bf Explanation of Columns:} 1=Source; 2=Net Chandra
exposure time in seconds after data screening; 3=Knot detected at X-rays;
4=Distance of X-ray knot from core in arcsec; 5=Position Angle of
X-ray knot in degrees; 6=Net X-ray counts in a 1\arcsec\ aperture in
0.4--8 keV. Uncertainties include the local background (see text). 


\noindent 
{\bf Notes:} a=Counterlobe; b=Uncertainty is 30\% the value of the count rates
(see text). 


\newpage 


\scriptsize
\begin{center}
\begin{tabular}{llrccrcc}
\multicolumn{8}{l}{{\bf Table 3: Jets broad-band emission}} \\
\multicolumn{8}{l}{   } \\ \hline
& & & & & & & \\
Source & Knot & F$_{5~GHz}$ & F$_{5852~\AA}$ & F$_{1~keV}$ & $\alpha_{ro}$ & $\alpha_{ox}$ & $\alpha_{rx}$ \\ 
& & & & & & & \\
(1) & (2) & (3) & (4) & (5) & (6) & (7) & (8) \\ 
& & & & & & &  \\ \hline
& & & & & & & \\
0723+679 & A & 73 & $<$ 0.06$^{a}$ & 0.4 & $>$ 0.90 & $<$ 0.81 & 1.07 \\ 
         & B & 110 & $<$ 0.06$^{a}$ & 1.1 & $>$ 0.92 & $<$ 0.65 & 1.04 \\
         & CL$^b$& 285& $<$ 2.8$^{a}$ & 0.24 & $>$ 1.0  & $<$ 0.90 & 1.18  \\
& & & & & & & \\
1055+018 & $\cdots^c$ & 69 & $<$ 0.06$^{a}$ & $<0.01^a$ & $>$ 2.27 & $\cdots$ &$>$1.28  \\ 
& & & & & & & \\
1136--135& A & 1 & 0.23 & 1.4 & 0.73 & 0.83 & 0.76 \\
         & B & 41 & 0.24 & 3.7 & 1.04 & 0.68 & 0.92 \\
         & C$^{~d}$ & 190 & 0.13 & $<$ 0.6$^a$ & 1.23 & $>$ 0.87 & $>$ 1.11 \\
         & CL$^b$& 520 & $<11$ &  $<0.01^a$ & $>$ 0.93 & $\cdots$ & $>$ 1.15  \\
& & & & & & & \\
1150+497 & A & 56 & 0.63 & 3.9 & 0.99 & 0.83 & 0.93 \\
         & B & 36 & 0.02 & 1.3 & 1.24 & 0.44 & 0.97 \\
         & C & 74 & 0.08 & 1.0 & 1.19 & 0.71 & 1.02 \\
         & CL$^b$& 45 & $<$ 7.0  & $<0.01^a$ & $>$ 0.76 &$\cdots$ & $>$ 1.06  \\
& & & & & & & \\
1354+195 & A & 57 & 0.3  & 8.2 & 1.05 & 0.60 & 0.90 \\
         & B & 23 & 0.04 & 0.24 & 1.15 & 0.68 & 0.98 \\
         & C & 13 & $<$0.06$^a$ & 0.37 & $>$1.07 &  $<$0.83 & 0.98 \\
         & D & 16 & $<0.06^a$  & 0.25 & $>$1.08 & $<$0.89  & 1.02 \\
         & E & 6  & $<0.06^a$ & 0.25 & $>$1.00 & $<$0.89 & 0.96 \\
         & F & 12 & $<0.06^a$ & 0.72 & $>$1.06  &$<$0.72 & 0.94 \\
         & G & 13 & $<0.06^a$ & 0.61 & $>$1.07 & $<$0.74 & 0.95 \\
         & H & $>$1 &$<0.06^a$ & 0.43 & $>$0.84 & $<$0.80 & 0.83 \\
         & I & 87 & $\cdots^e$ & 0.67 & $\cdots$ & $\cdots$& 1.06 \\
         & CL$^b$& 149 & $<$ 9.0  & $<0.01^a$ &$>$ 0.84 & $\cdots$ & $>$ 1.10 \\
& & & & & & & \\
2251+134 & Jet & 178$^f$ & $<$ 1.6 & $<0.01^a$  & $>$ 1.89 & $\cdots$  &$>$ 1.33 \\ 
& & & & & & &  \\ \hline

\end{tabular}
\end{center}

\scriptsize 
\noindent 
{\bf Explanation of Columns:} 1=IAU name of source; 2=Knot (see text); 3=Radio flux in mJy; 
4=Optical flux in $\mu$Jy; 5=X-ray flux in nJy; 6=Radio-to-optical spectral index;
7=Optical-to-X-ray spectral index; 8=Radio-to-X-ray spectral index. 


\noindent 
{\bf Notes:} a=3$\sigma$ upper limit; b=Counterlobe; c=Average from
two positions along the radio jet at 8.8\arcsec, PA=174 deg, and
6.2\arcsec, PA=175 deg; d=Radio/optical knot at 10\arcsec\ from the
core; e=Off the \hst\ field of view; f=Integrated on both radio lobes
(see Figure 2).


\newpage 


\scriptsize
\begin{center}
\begin{tabular}{llcrccrccc}
\multicolumn{10}{l}{{\bf Table 4: Parameters of Model Fitting to SEDs$^{~a}$}} \\
\multicolumn{10}{l}{   } \\ \hline
& & & & & & & & & \\
Source/Knot & Model & $\gamma_{max}$ & $n$ & $B$ & $K$ & $\delta$ & $\Gamma$ & P$_{jet}$ & P$_{rad}$\\
& & & & & & & & & \\
(1) & (2) & (3) & (4) & (5) & (6) & (7) & (8) & (9) & (10) \\ 
& & & & & & & & & \\ \hline
& & & & & & & & & \\
0723+679 A & CMB & $5 \times 10^4$ & 2.6 & 14 & 1.5 & 5 & 3 & 47.05 & 42.23 \\
~~~~~~~~~~~~~~B & CMB & $5 \times 10^4$ & 2.6 & 10 & 4.8 & 5 & 3 &   &  \\
& & & & & & & & & \\
1136--135 A & Sync & $5.7 \times 10^7$ & 2.5 & 1 & 0.1 & 7 & 5 & 46.47 & 43.12 \\
~~~~~~~~~~~~~~B & CMB & $3.5 \times 10^5$ & 2.6 & 4 & 1.0&7 & 5 & & \\
& & & & & & & & & \\
1150+497 A & CMB & $6 \times 10^5$ & 2.6 & 2.5 & 1.8 & 6 & 3.5 & 47.26 & 42.60 \\
~~~~~~~~~~~~~~B & CMB & $1 \times 10^5$ & 2.6 & 2.8 & 0.8 & 6 & 3.5 & & \\
& & & & & & & & & \\
1354+195 A & CMB & $2.5 \times 10^5$ & 2.6 & 3.6 & 5.7 & 6 & 4 & 47.88 & 43.70 \\
~~~~~~~~~~~~~~B & CMB & $2.5 \times 10^5$ & 2.6 & 5.6 & 0.8 & 6 & 4 & & \\
& & & & & & & & & \\ \hline

\end{tabular}
\end{center}


\noindent 
{\bf Explanation of Columns:} 1=Source and Knot; 2=Model used to model
the SED of the knot. Sync: synchrotron; CMB: Inverse Compton
scattering off the microwave background photons (see text); 3=Maximum
electron energy; 4=Index of the electron power law distribution
($N=N_0 \gamma^{-n}$); 5=Magnetic field (in $10^{-5}$ Gauss);
6=Normalization of the electron power law distribution in 10$^{-4}$
($N=N_0 \gamma^{-n}$); 7=Doppler factor; 8=Bulk Lorentz factor; 9=Log
of the intrinsic jet power (erg s$^{-1}$); 10=Log of the radiated jet
power (erg s$^{-1}$).

\vspace{0.1in}

\noindent{\bf Notes:} a=In all cases a region size $R=5 \times
10^{21}$ cm and minimum electron energy $\gamma_{min}=10$ was assumed.



\begin{references}

\reference {} Aharonian, F.A. 2001, MNRAS, submitted (astro-ph/0106037)

\reference {} Biretta, J.A. et al. 1991, AJ, 101, 1632

\reference {} Bridle, A.H. \& Perley, R.A. 1984, ARA\&A, 22, 319

\reference {} Brunetti, G., Setti, G., \& Comastri, A. 1997, A\&A,
325, 898 

\reference {} Brunetti G., Bondi M., Comastri A., \& Setti G. 2001,
A\&A, in press

\reference {} Celotti, A., Ghisellini, G., \& Chiaberge, M. 2001,
MNRAS, 321, L1

\reference {} Chartas, G. et al. 2000, ApJ, 542, 655

\reference {} Chartas, G., Gupta, V., Garmire, G., Jones, C., Falco,
E. E., Shapiro, I. I., \& Tavecchio, F. 2001, ApJ, in press
(astro-ph/0108277) 

\reference {} Cheung, C.C. et al. 2002, in prep. 

\reference {} Chiaberge, M., Celotti, A., Capetti, A. \& Ghisellini,
G. 2000, A\&A, 358, 104 

\reference {} Dermer, C. 1995, ApJ, 446, L63

\reference {} Feigelson, E.D. et al. 1981, ApJ, 251, 31

\reference {} Gambill, J.K. et al. 2002, in prep. 

\reference {} Gehrels, N. 1986, ApJ, 303, 336 

\reference {} Ghisellini, G. \& Celotti, A. 2001, A\&A, 379, L1 

\reference {} G{\' o}mez, J. 2001, Ap\&SS, 276, 281 

\reference {} Hardcastle, M.J., Birkinshaw, M., \& Worrall,
D. M. 2001, MNRAS, 326, 1499 

\reference {} Harris, D.E. 2001, in ``Particles and Fields in Radio
Galaxies'', R. A. Laing and K. M. Blundell, Editors, ASP Conference
Series, in press (astro-ph 0012374)

\reference {} Harris, D.E. \& Stern, C.P. 1987, ApJ, 313, 136


\reference {} Laing R. A., 1993, in Burgarella D., Livio M., O'Dea
C.P., eds Space Telescope Sci. Inst. Symp. 6: Astrophysical Jets.
Cambridge University press, Cambridge, p. 95

\reference {} Liu, F.K. \& Xie, G.Z. 1992, A\&AS, 95, 249 

\reference {} Murphy, D. W., Browne, I. W. A., \& Perley,
R. A. 1993, MNRAS, 264, 298 

\reference {} Owen, F.L. \& Puschell, J. J. 1984, AJ, 89, 932 

\reference {} Owen F. N., Hardee P. E., \& Cornwell T. J., 1989, 
ApJ 340, 698

\reference {} Pesce, J.E., Sambruna, R.M., Tavecchio, F., Maraschi,
L., Cheung, C.C., Urry, C.M., Scarpa, R. 2001, ApJ, 556, L79 

\reference {} Price, R., Gower, A.C., Hutchings, J.B., Talon, S.,
Duncan, D., \& Ross, G. 1993, ApJS, 86, 365


\reference {} Saikia, D.J., Shastri, P., Cornwell, T.J., 
Junor, W., \& Muxlow, T.W.B. 1989, Journal of Astrophysics and
Astronomy, 10, 203 

\reference {} Sambruna, R.M., et al. 2001, ApJ, 549, L161 

\reference {} Scarpa, R., Urry, C.M., Falomo, R., \& Treves, A. 1999, 
ApJ, 526, 643 

\reference {} Scarpa, R. et al. 2002, in prep. 

\reference {} Schwartz, D.A. et al. 2000, ApJ, 540, L69

\reference {} Sikora, M., Madejski, G., Moderski, R., \& Poutanen, J.\
1997, ApJ, 484, 108

\reference {} Sparks, W.B., Biretta, J.A., \& Macchetto, F. 1994, ApJS, 
90, 909 

\reference {} Tavecchio, F., Maraschi, L., Sambruna, R.M., \& Urry,
C.M. 2000, ApJ, 544, L23

\reference {} Thompson, A.R., Clark, B.G., Wade, C.M., \& Napier,
P.J.\ 1980, ApJS, 44, 151

\reference {} Urry, C.M. \& Padovani, P. 1995, PASP, 107, 803 

\reference {} Wardle, J.F.C. \& Aaron, S.E. 1997, MNRAS, 286, 425 

\reference {} Wilson, A.S., Young, A.J., \& Shopbell, P.L. 2001, ApJ,
547, 740 

\reference {} Worrall, D.M., Birkinshaw, M. \& Hardcastle, M.J. 2001,
MNRAS, 326, L7  

\end{references}
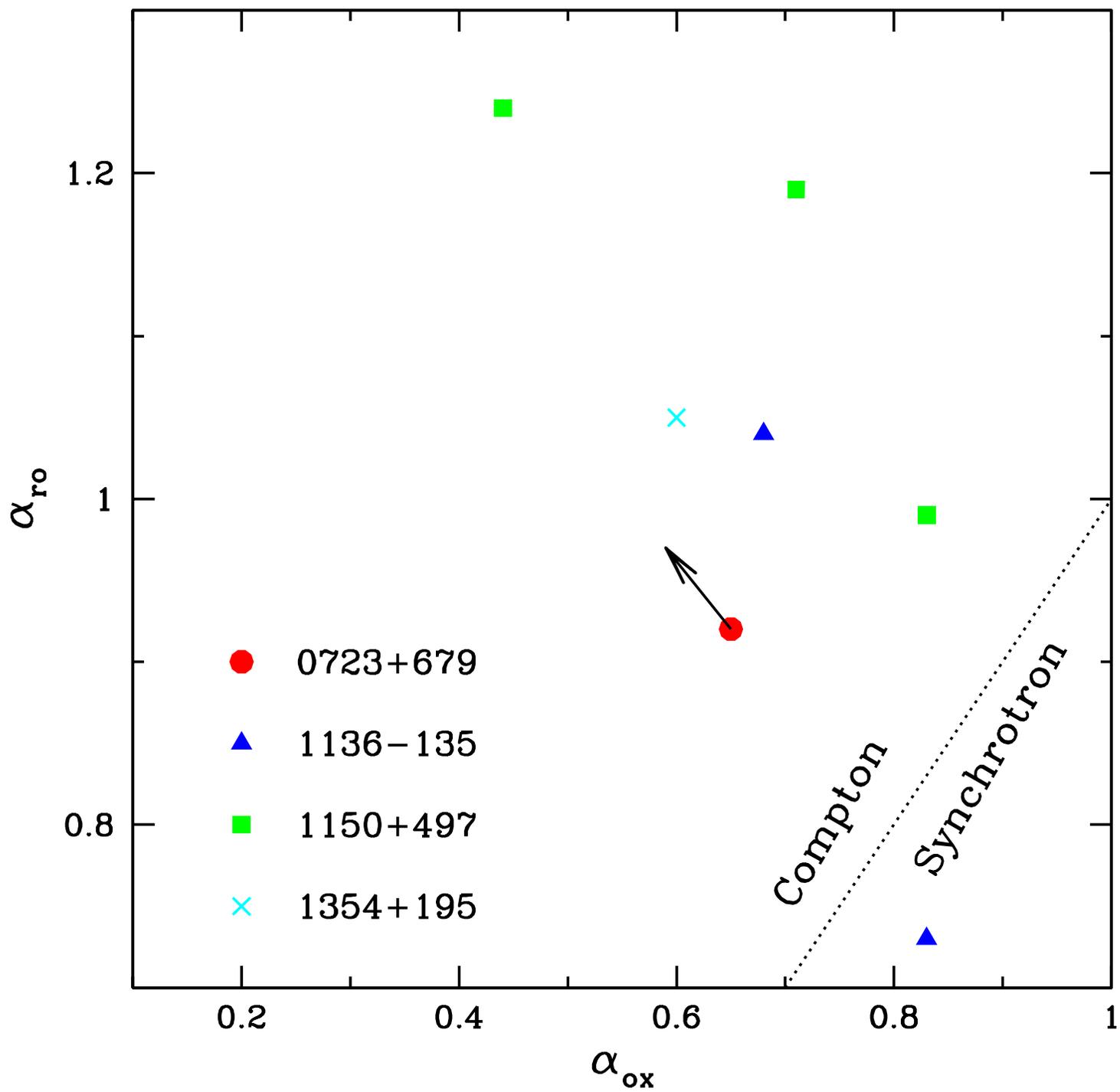
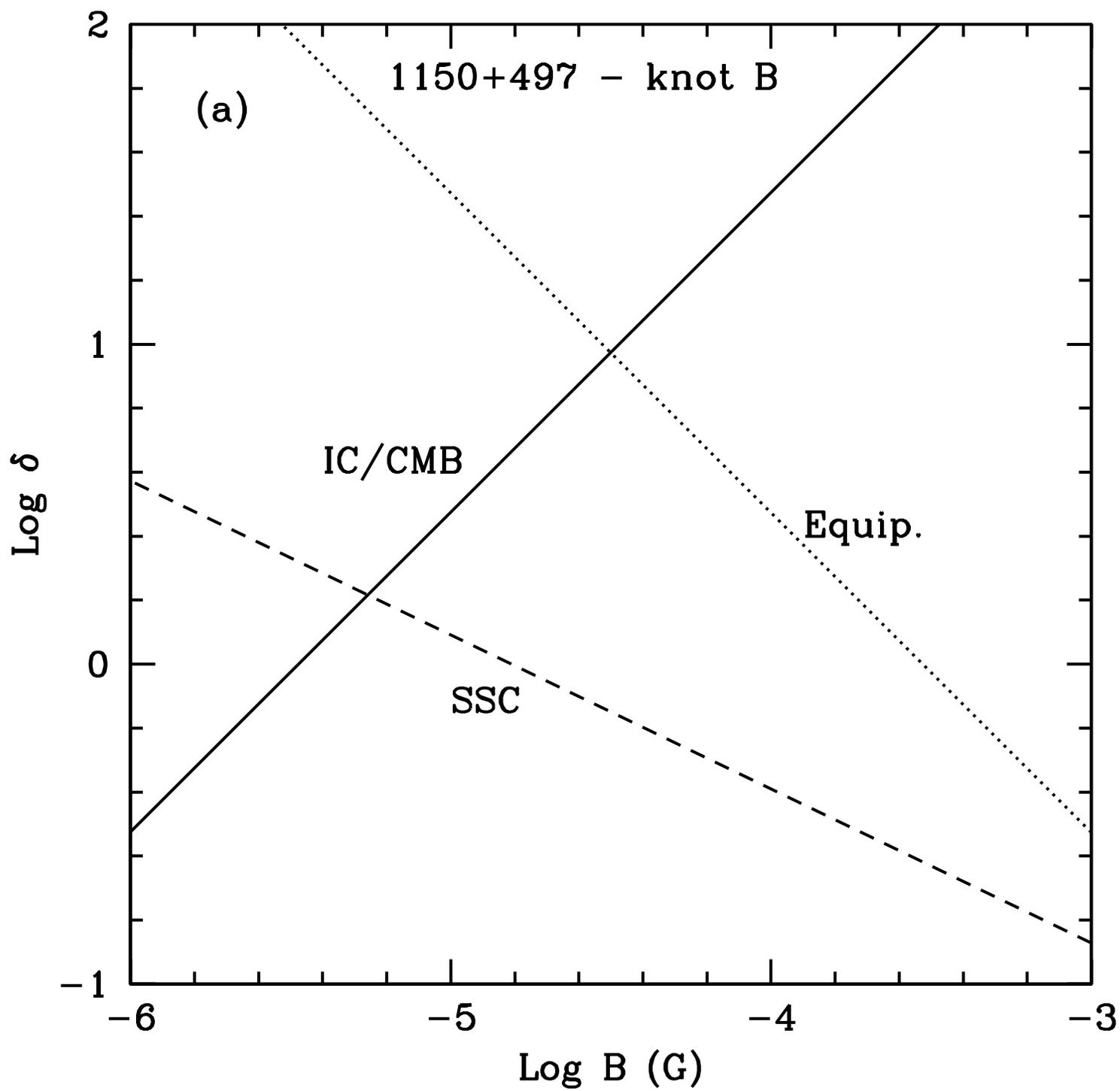
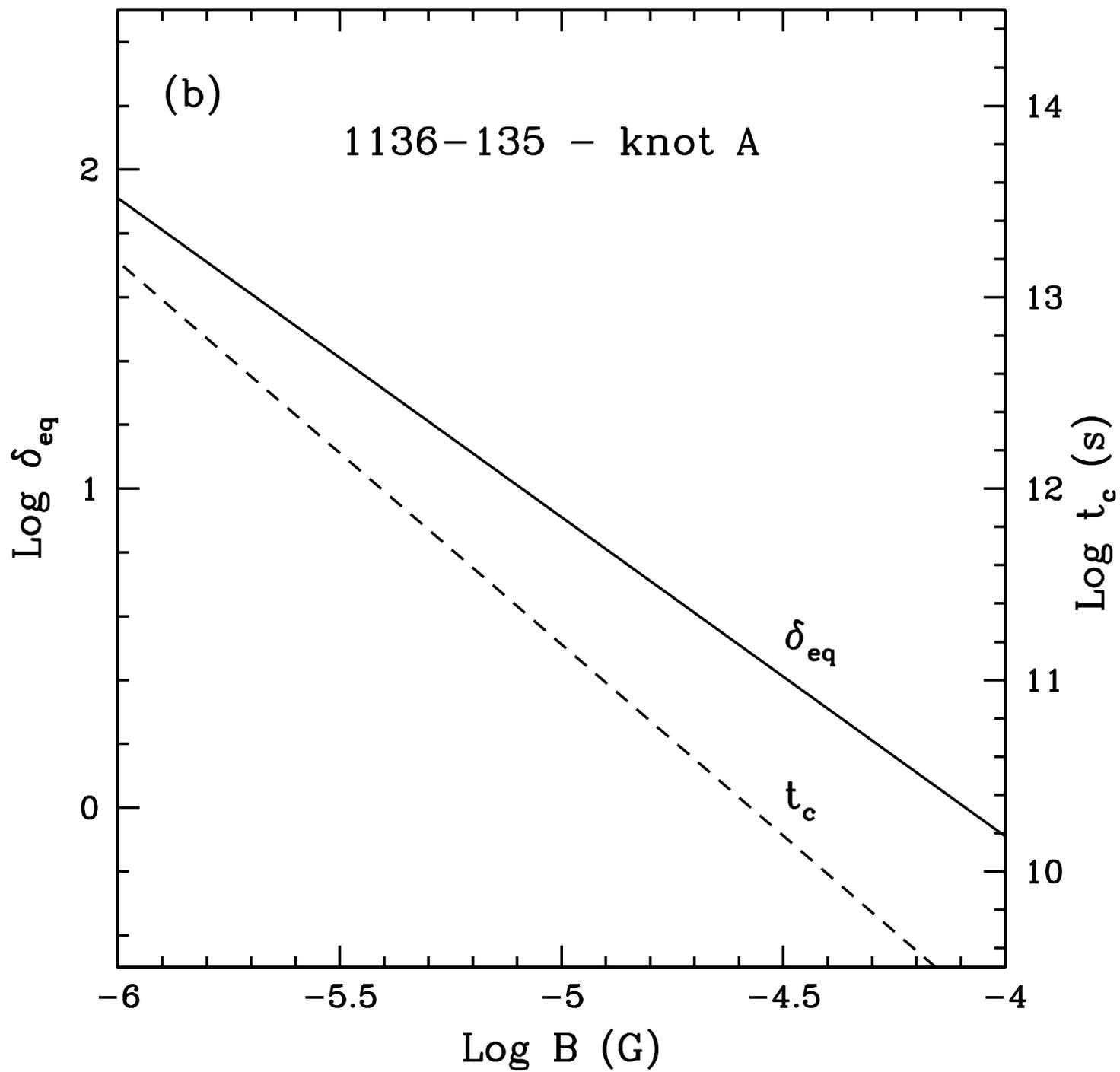
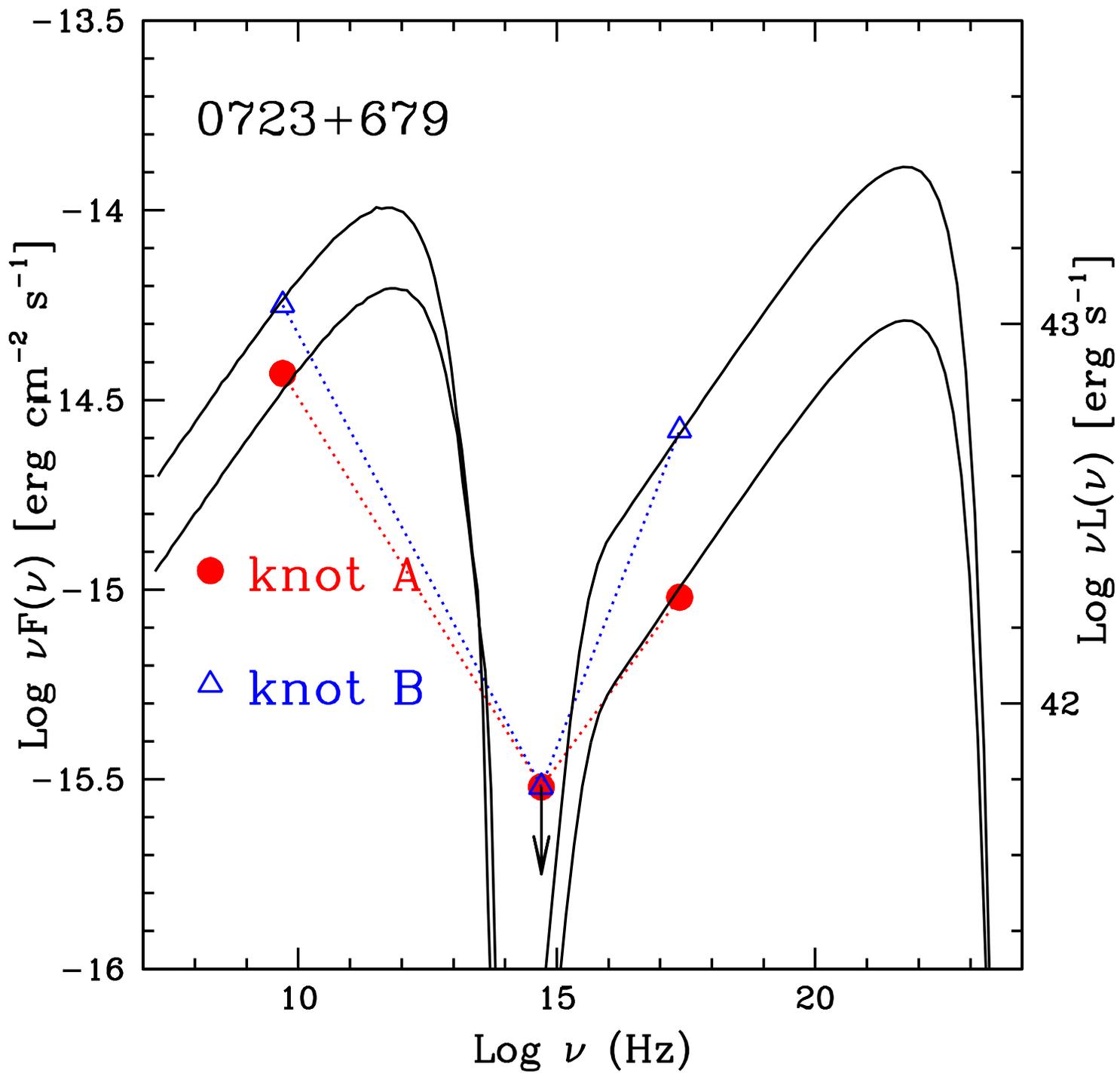
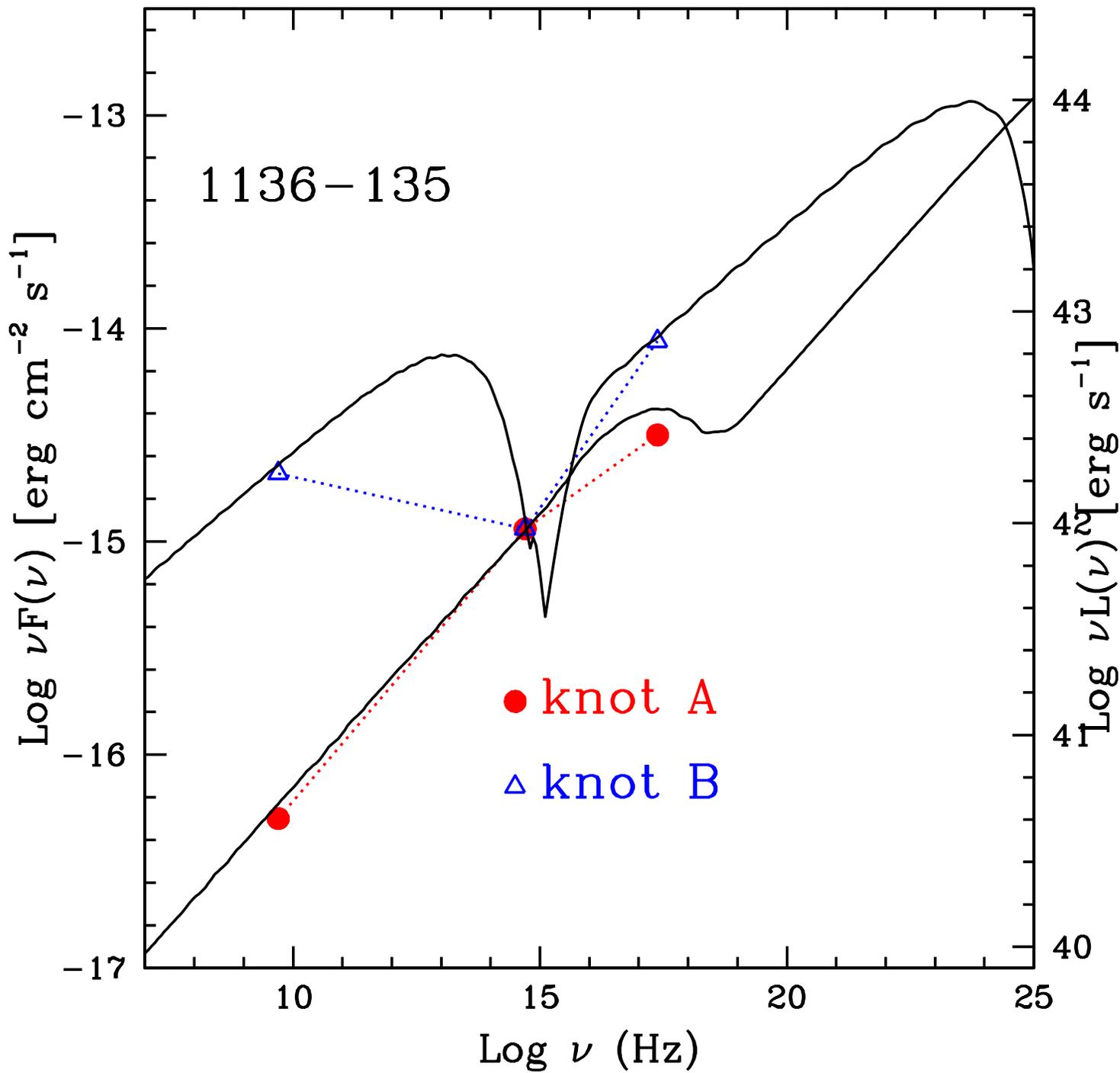
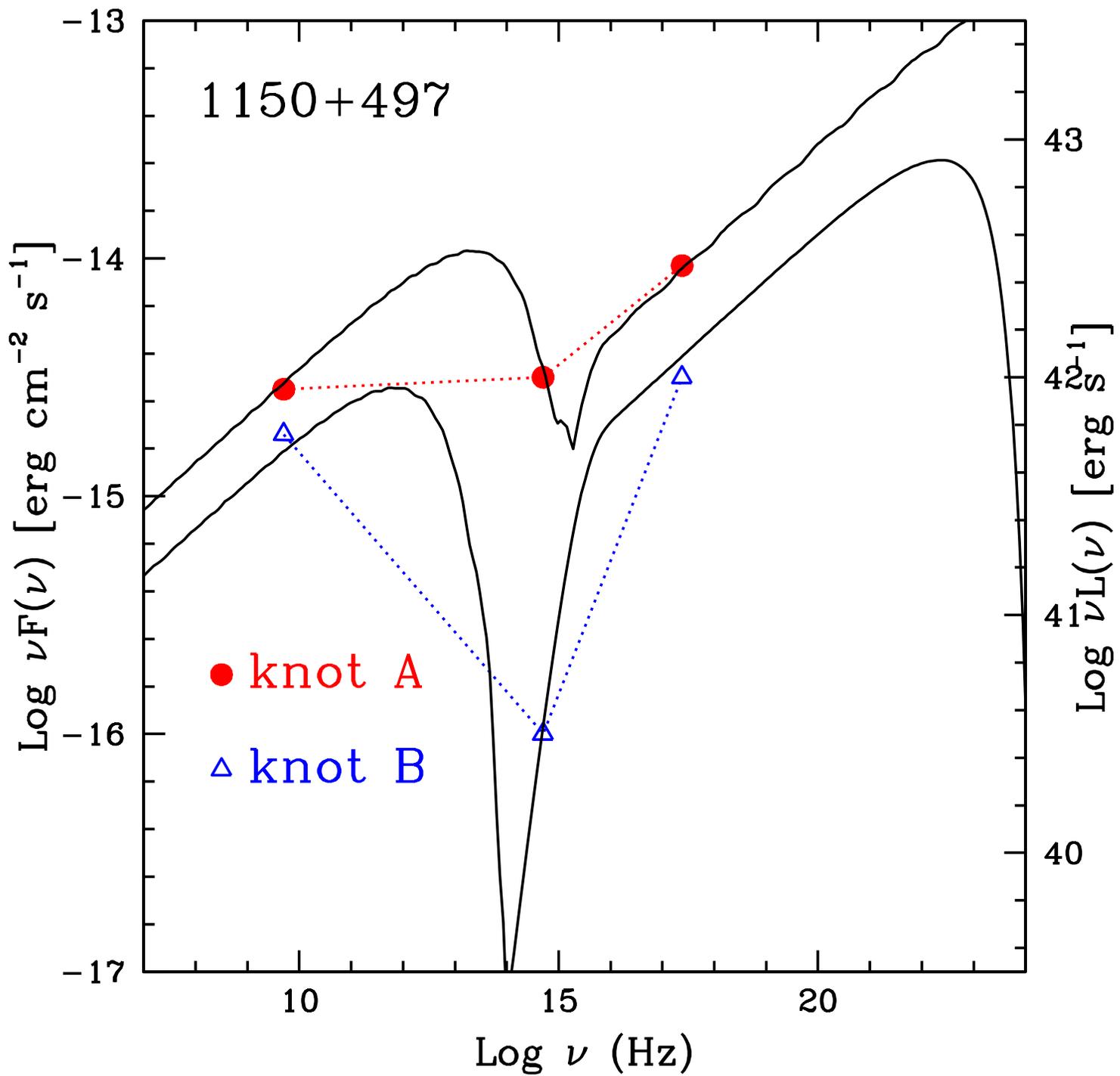
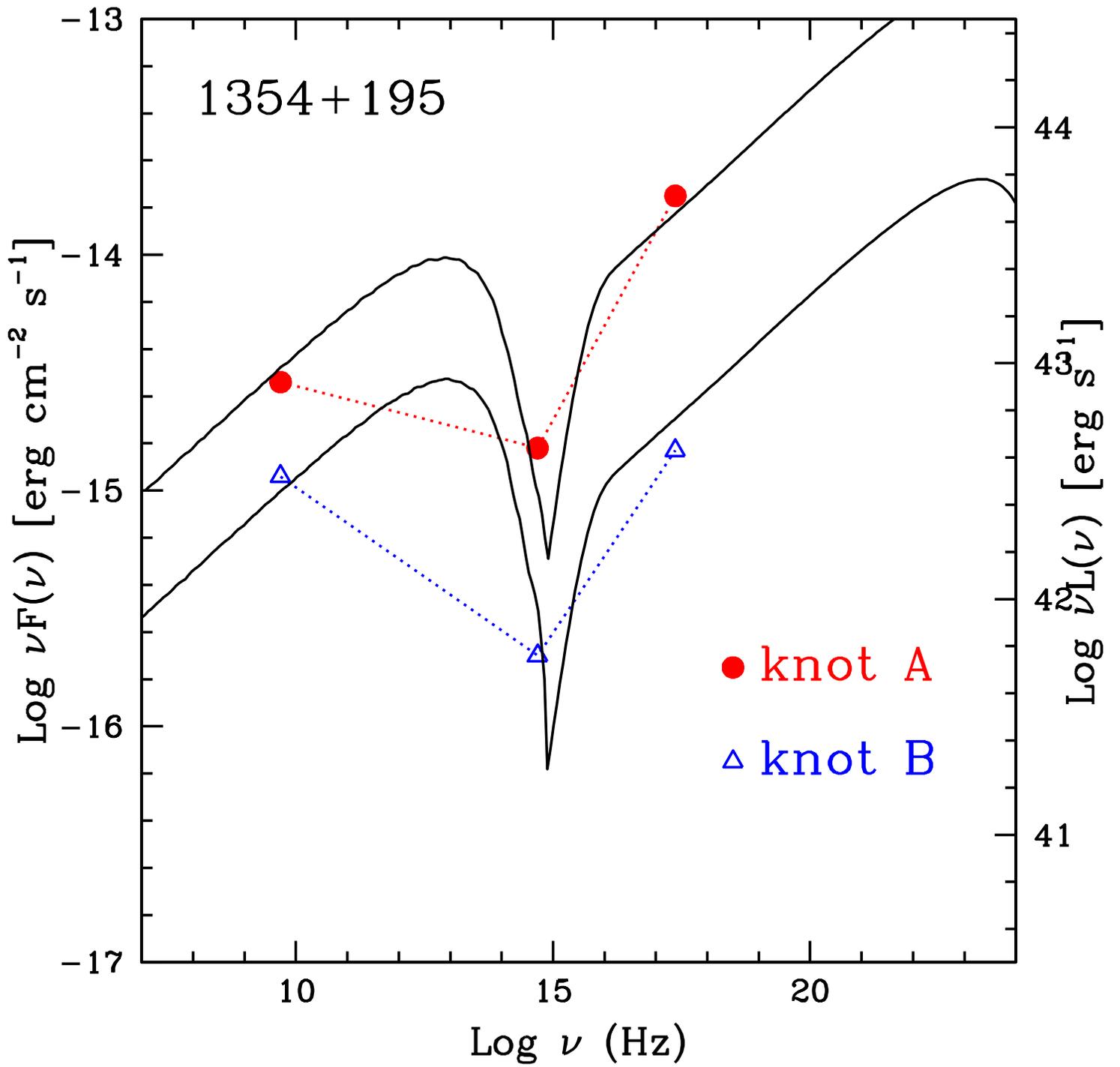
\end{document}